\documentclass[aps,pra,twocolumn,nofootinbib]{revtex4-1}
\usepackage{graphicx}% Include figure files
\usepackage{dcolumn}% Align table columns on decimal point
\usepackage{bm}% bold math
\usepackage{hyperref}% add hypertext capabilities

\usepackage{verbatim}
\usepackage{rotating}
\usepackage{array}
\usepackage{amssymb}
\usepackage{tikz}

\usepackage{amsmath}
\usepackage{amssymb}

\def\v#1{{\bf#1}}
\def\be{\begin{equation}}
\def\ee{\end{equation}}
\def\bea{\begin{eqnarray}}
\def\eea{\end{eqnarray}}

\def\<{\langle}
\def\>{\rangle}

\newcommand{\Tr}[1]{{\textrm{Tr}}{\left\{#1\right\}}}

\newcommand{\ket}[1]{|{#1}\rangle}
\newcommand{\bra}[1]{\langle{#1}|}
\newcommand{\ketbra}[2]{|{#1}\rangle\langle{#2}|}
\newcommand{\braket}[2]{\langle{#1}|{#2}\rangle}

\begin{document}
%\nocite{*}
%\preprint{APS/123-QED}

%\title{Wigner function in the dressed-states phase space}
\title{Wigner function in the polariton phase space}
%\title{Wigner function in an hybrid phase  space}

\author{A. Rosado} 
\author{E. Sadurn\'i} 
\author{J. M. Torres} 
\email{jmtorres@ifuap.buap.mx}
\affiliation{Instituto de F\'isica, Benem\'erita Universidad Aut\'onoma de Puebla, Apdo. Postal J-48, Puebla, Pue. 72570, M\'exico}
%\affiliation{Instituto de F\'isica, Benem\'erita Universidad Aut\'onoma de Puebla, Apdo. Postal J-48, Puebla, M\'exico}

\begin{abstract}
The Wigner function of a dynamical infinite dimensional lattice is studied.  A closed differential equation without diffusion terms for this function is obtained and solved. We map atom-photon interaction systems, such as the Jaynes-Cummings model,  into this lattice model, where each dressed or polariton state corresponds to a point in the lattice and the conjugate momenta are described by the eigenvalues of the phase operator. The corresponding Wigner function is defined by these two conjugate variables in what we name the polariton phase space. We derive a general propagator of the Wigner function, which is also valid for other hybrid models. 
\end{abstract}


%\pacs{03.65.Fd, 02.30.Tb, 05.45.Df }

%\keywords{diffractive focusing, discrete models}
%\submitto{\IP}

\maketitle

\section{Introduction}
\label{Intro}
In one of his books \cite{feynman}, R. P. Feynman gives an account of a discussion with his father about the nature of emitted light: ``Is the photon in the atom ahead of time?... where does it come from? How does it come out?'' There is an obvious answer to this question, regarding the way radiation is produced by the motion of charges even at the level of classical physics. But when it comes to photons --a quantum-mechanical notion-- this naive question shows the importance of formulating statements about nature in terms of quantities that can be measured. In the quantum world, we have to measure first the (emitted) photon in order to infer
that it was produced. 
Therefore the number of particles is, above all, a quantum observable. 
Interestingly, we have at least two layers of this discussion that drive our curiosity in the present paper:

(a)~For the Jaynes Cummings model of atom-radiation interaction (used as an approximation) Feynman gave the right answer  in the sense that the photon is not only produced, but its existence as a particle was caused by the de-excitation of the atom, i.e. one excitation, one photon. This entails the conservation of excitations in the whole quantum system.  
(b)~For the Rabi model of the atom plus radiation (without approximations) the absorption or production of photons is not necessarily in one-to-one correspondence with one jump of the electron upwards or downwards, respectively. Although energy is conserved, the excitation number is not. One has to measure with a photocounter and record the average number of emitted particles and their fluctuations around it; this is the full answer.

Our aim in this work is not to delve more into philosophical matters, but to provide a visualization of the phenomenon stated  above, by employing suitable mathematical tools such as a two-dimensional quantum phase space containing the eigenvalue of the phase operator and the number of atomic excitations as variables. This shall be done via quasi-distributions. Several numerical studies of the Rabi model have been carefully carried out \cite{braakspecial} and even a proof for the integrability of the Rabi model has been reached 
\cite{braak}, but since no analytical solutions are known beyond the rotating wave approximation, a simple description of excitation fluctuations seems to be lacking.

The Wigner function is the tool of choice when it comes to the quantum dynamical description of oscillatory systems in phase space 
that has been mostly  employed with continuous variables such as position and momentum, or time and frequency. 
Interestingly, conjugate variables can be found also in lattices \cite{levi,sadurni2014,hernandez-espinosa}, where it is advantageous to employ operator methods in order to elucidate dynamical features of tight-binding models \cite{sadurni2013,hartmann}. Moreover, peculiar connections \cite{villafane,sadurni2010} between a discrete position operator and the number of quanta of relativistic oscillators motivate further the definition of Wigner functions in lattices made of excitations.
These objects have been previously considered for discrete finite systems, for instance in Refs. \cite{wootters1,miquel,wootters,arguelles}. 
A number-phase Wigner function has also been
investigated for the harmonic oscillator in the frame of quantum optics \cite{moya,carvalho}. 
Furthermore, the problem has been investigated in the case of a periodic continuous variable \cite{berry,bizarro}, 
where the corresponding conjugate variable is infinite but discrete. 
In \cite{hinarejos} different types of states in a lattice and their 
corresponding phase space were studied, but without taking into account any dynamical process.

In this paper we consider a generalization of Wigner's function to lattices and study the dynamics
of two different types of physical models that can be mapped into this lattice. 
The first type is a tight-binding array modelling photonic crystals, where the position 
indicates the location of a lattice cell. 
This can also be connected to spin chains in a one-excitation manifold. 
The second class of systems which can be mapped onto this lattice are atom-photon interaction models.
The most prominent example is a two-level atom modelled by a Jaynes-Cummings Hamiltonian, 
where the atom-photon dressed or polariton states \cite{hartmann2006,angelakis,restrepo,koch} can be connected to 
lattice sites with a conjugate periodic continuous variable.
The phase space obtained by these two variables will be called 
{\it polariton phase space}. We derive a general propagator of the Wigner function for 
systems with discrete spectrum, which can also be applied to other atom-photon interaction models,
some of which are also discussed in this work. 
Regarding the mathematical developments of this paper, we show that closed differential equations for the evolution of discrete Wigner functions can be obtained. This allows to infer the shape of a quasi distribution by means of a propagator in phase space. In our view, this is an important extension to the very few known examples where phase space evolves in a simple manner (e.g. bilinear Hamiltonians in $x,p$); a particular algebraic structure is behind it, but in our case it is related to the discrete translation group, rather than the dynamical algebra of an oscillator. The wider set of operations in ISO$(n)$ was studied in \cite{wolf,atakishiyev}, but to our knowledge no connection with discrete Wigner functions has been provided.

The paper is organized as follows. In Sec. \ref{DiscreteWigner}, we introduce a discrete Wigner function and describe
its properties.  In Sec. \ref{Half} we discuss the requirements that have to be fulfilled by the Wigner function
and what this implies for the semi-discrete phase space. 
In Sec. \ref{Tight}, we consider the dynamics of a simple lattice model which can also be used
to study transport in spin chains. Sec. \ref{Propagator} is devoted to the study of systems with discrete 
spectrum and a general propagator for these type of system is introduced. Photon atom models
are mapped and discussed in Sec. \ref{Polariton}, where we consider the Jaynes-Cummings and the Buck-Sukumar 
models. We present our conclusions in Sec. \ref{Conclusion}. 

\section{The discrete Wigner function}
\label{DiscreteWigner}
A straightforward adaptation of the Wigner function \cite{case,wigner} for a 1D lattice
can be obtained by replacing from the original definition the
Fourier transform with its discrete analogue. 
In the following, we shall  elucidate that although there are a number of ways of doing this, all of them can be reduced to the same equivalent principle of correct marginal distributions. Either a half integer lattice or doubling phase space \cite{hinarejos,miquel,carvalho}, or an intrinsic spinor structure \cite{klimov} can be induced in discrete phase space to show the validity of this object.  A formulation of the Wigner function in discrete position and periodic momentum space can be written as
\begin{equation}
  W(n,k) \equiv \frac{1}{2\pi}\sum_{n'=-\infty}^{\infty} \psi^*_{n-n'} \psi_{n+n'} e^{-2ikn'}, 
\label{1}
\end{equation}
where the functions in space and Bloch quasi momentum $k$ are such that
\bea
\psi_n = \< n | \psi \>, \quad \< n | k \> = \frac{1}{\sqrt{2\pi}} e^{ikn},
\label{2}
\eea
with $n \in \mathbb{Z}$ and  $k \in [0,2\pi]$.
Accordingly, the position and phase operators can be represented by
\begin{equation}
 N |n\> = n |n\>, \quad T |k\> = e^{-ik} |k\>, 
\label{3}
\end{equation}
and it can be shown that they fulfill the following relations
\begin{equation}
T|n\>=|n+1\>, \quad \left[ N , T\right]=T, \quad TT^{\dagger}=T^{\dagger}T= \v 1.
  \label{relT}
\end{equation}
For later convenience, we also introduce an inversion operator with the properties
\bea
 I \equiv \sum_{n=-\infty}^{\infty} |- n \>\<n|, \quad \left\{ N, I \right\} = 0, \quad T I = I T^{\dagger}. 
\label{3.1}
\eea
After performing position summations or first Brillouin zone integrations, we can see that the definition (\ref{1}) reduces to probability distributions in the corresponding conjugate variable.
In doing this, one is led to the conclusion that $n$ and $n'$ should be allowed to take half-integer values, this
will be discussed in more detail in Sec. \ref{Half}.

We introduce a notation that allows to conceive $W$ as the quantum average of an operator, namely
\bea
\hat W(n,k)\equiv \frac{1}{2\pi} e^{ik N}T^{2n}I  e^{-ik N}, 
\label{4}
\eea
from which a  simple relation follows
\bea
W(n,k) = \< \psi | \hat W(n, k)  | \psi \> = \< \hat W \>_{\psi},
\label{5}
\eea
or $\Tr{\rho\hat W}$ for a general mixed state $\rho$.
As an alternative and equivalent definition we introduce 
\begin{equation}
  \hat W(n,k)=\frac{1}{2\pi}\sum_{n'=-\infty}^\infty e^{-i2 k n'}|n-n'\>\<n+n'|,\quad 
  \label{Woperator}
\end{equation}
where $n,n'\in\frac{1}{2}\mathbb{Z}$, but restricted to  
$n\pm n'\in \mathbb{Z}$, and $k \in [0,2\pi)$. The inclusion of half-integers will 
be discussed in the next section.

Usually, the evolution of the Wigner function is computed by considering $|\psi(t)\>=U_t |\psi(0)\>$, but in the notation above, we may study the evolution of this object by fixing $\psi$ or $\rho$ and letting $\hat W$ vary in time, i.e. in 
the Heisenberg picture. We have the following differential equation for the operator in the Heiseberg picture ($\hbar =1$)
\begin{equation}
i \frac{\partial \hat W(n,k;t)}{\partial t} = \left[ \hat W(n,k;t) , H \right] 
\label{6}
\end{equation}
or equivalently
\begin{align}
  \hat W(n,k;t) &= U^{\dagger}(t) \hat W(n,k)U(t) 
  \nonumber\\
  &=\frac{1}{2\pi} 
e^{ik N(t)}T^{2n}(t)I(t)  e^{-ik N(t)}. 
\label{7}
\end{align}
Note that in \eqref{6} the total derivative with respect to time appears on the l.h.s., but for notational convenience and since
neither $n$ nor $k$ depend on $t$, we replace it with a partial derivative with respect to $t$.
Note how the use of the Moyal-Poisson bracket has been avoided in this notation.
Sometimes, it is better to employ the explicit forms of $N(t), T(t)$ if the equations of motion allow exact solutions.

\section{The ``half-excitation'' problem}
\label{Half}
In this section we consider the problem arising from the doubling  phase space points
for discrete Wigner functions. This issue has its root in the definition \eqref{1} or \eqref{Woperator}, where
the summation is carried out for terms of the form $n\pm n'$. If one considers a 
bidimiensional infinite space labelled by $n$ and $n'$, the 
former corresponds to summing in diagonals rotated $45^\circ$ from the original lattice. 
The effect is well known by the community, for instance in \cite{arguelles} the authors recognized this
lattice as a ``based-centered cubic'' whose size had to be doubled to become ``simple-cubic''.
In \cite{hinarejos} the lattice was reindexed in order to account for the doubling of
points, but  retaining an integer value. 
In the following we first comment on what we 
consider the most straightforward  way to solve this issue which is to allow half integer values
for $n$ and $n'$ in \eqref{1}. 
We refer to this effect as the ``half-excitation'' problem.
Then, we provide a more general construction which augments
the definition of the Wigner function to a spinor matrix, circumventing  the use of half-integers. 
The reader might skip this section without loss of continuity, as in the rest of the paper
we  consider the definition in \eqref{1} for integer lattice points and with quasi-momentum coordinate in the reduced interval 
$[0,\pi]$.

\subsection{The half-integer lattice}
The need for half-integers in phase space becomes more evident when evaluating
the marginals of the Wigner function, which should give the probability density distribution. 
Integrating over the generalized momentum we obtain the position density distribution as
\begin{align*}
  \int dk W(n,k)&=\sum_{n'}\int_0^{2\pi} 
  \frac{dke^{-2ikn'}}{2\pi}\bra{n+n'}\rho\ket{n-n'}
\nonumber\\
&=\bra{n}\rho\ket{n}.
%  \label{}
\end{align*}
One can note that there is no redundancy in the domain of integration 
if the summation is taken over $n'\in\frac{1}{2}\mathbb{Z}$.
Using the kernel $e^{2imk}$ in the previous integral,  
one can also retrieve all matrix elements of the form $\bra{n+m}\rho\ket{n-m}$, where
it is evident that elements of the form $\bra{n}\rho\ket{n+1}$ require 
half integer values to be allowed, as pointed out in \cite{carvalho}.
Summing over all lattice points we should get the momentum distribution as 
\begin{equation*}
  \sum_{n=-\infty}^\infty W(n,k)=\sum_{n,n'=-\infty}^\infty\frac{1}{2\pi}  
  e^{-2ikn'}\bra{n+n'}\rho\ket{n-n'}.
\end{equation*}
Once more, the sum over $n$ and $n'$ has to be carried out over half integers, i.e., $n,n'\in\frac{1}{2}\mathbb{Z}$,
but restricted to $n+n'\in\mathbb{Z}$.
The argument in the exponential can be rewritten as $2kn'=k(n+n')-k(n-n')$. By defining
$m=n+n'$ and $m'=n-n'$ and considering the overlap between position and momentum eigenstates
in Eq. \eqref{2}, one gets
\begin{align*}
  \sum_{n=-\infty}^\infty W(n,k)&=\sum_{m,m'}
  \braket{k}{m}\bra{m}\rho\ket{m'}\braket{m'}{k}
  =\bra{k}\rho\ket{k},
%  \label{}
\end{align*}
where we have used the completeness relation of the position basis in the last step. Note that here it is
important that the sum over $n$ runs over half integer values, otherwise $m$ would run only
over even integers and one would not obtain the correct marginal distribution.  

Another important and useful property is the completeness of the operators defined in 
\eqref{Woperator} which allows to express any Hermitian operator $O$ with the following expansion
\begin{align}
  O&=
  2\pi\sum_{n=-\infty}^\infty\int_0^{2\pi} dk \Tr{O\hat W(n,k)}\hat W(n,k)
  \nonumber\\
  &=\sum_{n,n'}\bra{n-n'}O\ket{n+n'}\ketbra{n-n'}{n+n'}.
  \label{complete}
\end{align}
One can easily check that the above expression is valid when considering half-integer
values of $n$ and $n'$. 
It should be stressed however, that although $n$ in $W(n,k)$ can take half-integer values,
the values of $n\pm n'$ in \eqref{Woperator} have to be  always integer as they label
position states that are undefined  for half-position or half-excitation.

\subsection{The Wigner function  as a spinor matrix}
It is possible to define a Wigner function with the correct marginal distributions without 
introducing a semi integer lattice, or without doubling the position value as in \cite{hinarejos}.
In case of atomic excitations, such half-integer values are important for coding the coherences of the density matrix, but they also lead to a more abstract interpretation of phase space with positions that stand for 
``half a photon''; let us solve this problem. We are convinced that our straightforward definition in \eqref{1} is already valid for distributions that occupy exclusively even or odd sites on the discrete line. This is not the case for general wavefunctions, so let us extend our definition to a $2 \times 2$ array of phase space planes. Since the group structure associated with the lattice allows the construction $\mathbb{Z} \cong 2\mathbb{Z} \otimes \mathbb{Z}_2$, even integers are a normal subgroup of $\mathbb{Z}$ and the remaining part is a trivial binary sum. As a warm up, let us focus on $\mathbb{Z}_2$ alone. Consider a function $\psi$ of only two sites with values $\psi_0$, $\psi_1$ and its finite Fourier transform onto the functions of well-defined parity
\bea
\tilde \psi_{\pm} = \frac{1}{\sqrt{2}} \left( \psi_0 \pm \psi_1 \right). 
\label{w1}
\eea
The Wigner function $W(s,p)$ has four possible values according to $s=0,1$ and $p=-,+$. The most general expression that fulfills the right marginals
\bea
\sum_{s=0,1} W(s,p) = |\tilde \psi_p|^2, \qquad  \sum_{p=\pm} W(s,p) = |\psi_s|^2,
\label{w2}
\eea
is the set of sums \footnote{This definition does not have the form of a convolution. Note also that other definitions for spin $1/2$ are plausible, in  particular see \cite{klimov,agarwal}.}
\begin{align}
  W(0,+)&= A |\psi_{0}|^2 + B |\psi_1|^2 + 2C {\rm Re}\left[ \psi_0 \psi_1^* \right], \nonumber \\ 
  W(0,-)&= (1-A) |\psi_{0}|^2 - B |\psi_1|^2 - 2C {\rm Re}\left[\psi_0 \psi_1^* \right], 
  \nonumber \\
  W(1,+)&= \frac{|\psi_0+\psi_1|^2}{2}-W(0,+) 
  %\nonumber \\
  %W(1,+)&= \frac{1-2A}{2} |\psi_{0}|^2 + \frac{1-2B}{2} |\psi_1|^2 
  %\nonumber\\&
  %+ (1-2C) {\rm Re}\left[ \psi_0 \psi_1^* \right], 
  \nonumber \\
  W(1,-)&=|\psi_1|^2-W(1,+)
\label{w3}
\end{align}
where $A,B,C$ are arbitrary real numbers. With appropriate definitions, these relations can be written as
\bea
W(l,\beta) = \sum_{a=0,1; b=0,1} C_{a,b}^{l,\beta} \psi_{a} \psi_b^{*}.
\label{w4}
\eea
In analogy with this treatment, the definition for $\mathbb{Z}$ requires four rows of sums. With two indices $s=\pm, p=\pm$ we get the needed structure if
\begin{align}
  W^{sp}(m,k) & \equiv  \frac{A^{sp}}{2\pi} \sum_{n=-\infty}^{\infty} \psi_{m+n} \psi^*_{m-n} e^{-2ikn} 
 \nonumber \\ 
 &+ \left( \frac{B^{sp}}{2\pi} \sum_{n=-\infty}^{\infty} \psi_{m+1+n} \psi^*_{m-n} e^{-2ikn} +  \rm{c.c} \right)
 \nonumber \\ 
 &+ \frac{C^{sp}}{2\pi} \sum_{n=-\infty}^{\infty} \psi_{m+1+n} \psi^*_{m+1-n} e^{-2ikn} 
\label{w5}
\end{align}
where now all terms with even and odd differences are considered in the sums, such as $\psi_{\mu}\psi^*_{\mu -2n}$ and $\psi_{\mu}\psi^*_{\mu -2n-1}$. The coefficients $A, B, C$ are fixed by the following marginals
\bea
\sum_{s=\pm} \int_{0}^{2\pi} dk W^{sp}(m,k) =
\begin{cases}
  |\psi_m|^2, & p = +\\
  |\psi_{m+1}|^2, & p = -\\
\end{cases}
%\cases{|\psi_m|^2 & for $p = +$\\
%|\psi_{m+1}|^2 & for $p = -$\\}
\label{w6}
\eea
and
\bea
\sum_{m=-\infty}^{\infty} \sum_{p=\pm} W^{sp}(m,k) =
\begin{cases}
a \, |\tilde \psi_k|^2, &  s = +\\
d\,|\tilde \psi_{k+\pi}|^2, & s = -\\
\end{cases}
\label{w7}
\eea
with $a,d>0$ normalization constants. A straightforward computation of 
(\ref{w6}) and (\ref{w7}) using (\ref{w5}) leads to the following consistency relations 
\begin{align}
&a=d/8, \quad A^{++} + C^{++} = A^{--} + C^{--},  
\nonumber \\ 
&C^{-+} + C^{++} = A^{+-} + A^{--}=0,\quad C^{+-} + C^{--}=1 
\nonumber \\ 
&B^{++} + B^{-+} = B^{--} + B^{+-}=0, \quad A^{++} + A^{-+}=1, 
\nonumber \\  
&B^{++} =e^{-ik}b,\quad B^{--} = e^{-ik}c, \quad b-c=1/2.
\label{w8}
\end{align}
This is the desired result. From the initially free 12 coefficients we are left with 4 degrees of freedom. See the appendix \ref{ApComp} for a full computation of marginals.
As can be noted, the general use of operators 
\bea
\hat W(n,k)= \frac{1}{2\pi} e^{ik N}T^{n}I T^{-n\pm 1} e^{-ik N}, 
\label{opWsim}
\eea
will throw us back to the dynamics of \eqref{4}. In the rest of the paper we shall only use the definition in \eqref{4} without the half-integer lattice and with quasi-momentum coordinate in the reduced interval 
$[0,\pi]$.

\section{The tight binding chain: A paradigmatic discrete model}
\label{Tight}
%Tight binding models are 
A relevant example well described by this model is an $XX$-spin chain  in the one-excitation submanifold and in the limit of infinite number
of sites. However, the description for a finite number of sites can be given also by  a truncation
of the Wigner function considered here. Although our aim in this work is to address atom-field models, we would
like to point out that this type of formalism is also useful to study transport in crystals
\cite{bloch,russel,longhi,zeilinger}, including spin systems \cite{batchelor}. 
Furthermore, any tight binding model is susceptible of this treatment, including Anderson
models with uncorrelated disorder as well as Harper's model with quasi periodic 
potentials \cite{roati}.
Let us investigate the convenience of the formalism with a specific example.

If the Hamiltonian is chosen as a nearest-neighbor homogeneous chain of sites with position operator $N$ and unit couplings, in the appropriate phase convention we may write
\begin{equation}
H = i T - i T^{\dagger}.
\label{8}
\end{equation}
The equation of motion (\ref{6}) becomes
\begin{align}
i \frac{\partial \hat W }{\partial t} &= \hat W i(T - T^{\dagger}) -i(T - T^{\dagger})\hat W  
%\nonumber \\
%&=& i(T^{\dagger} e^{-2ik} - T e^{2ik}) \hat W-  i(T - T^{\dagger})\hat W   
\nonumber \\
%&=& 
&=- 2i \cos(k) ( T e^{ik} - T^{\dagger} e^{-ik} ) \hat W
\label{9}
\end{align}
where we have used the properties of translation (\ref{3}) and inversion operators (\ref{3.1}). Note that this result still involves operations acting on $\hat W$ that do not resemble $\hat W$ again. We show now that the second derivative leads to a closed result:
\begin{align}
 \frac{\partial^2 \hat W(n, k)}{\partial t^2} &= 
4 \cos^2(k)\,\times 
\label{12}
\\
&\left[ \hat W(n+1,k) + \hat W(n-1,k) - 2 \hat W(n,k)\right].
\nonumber
\end{align}
where we have used 
the identities
\begin{align}
 T \hat W(n,k) T^{\dagger}&= \hat W(n+1,k),\,\,\, 
T \hat W(n,k) Te^{2ik}=\hat W(n,k), 
 \nonumber\\
 T^{\dagger} \hat W(n,k) T&= \hat W(n-1,k), 
\label{11}
\end{align}
that follow from their definitions. %

The result in Eq. \ref{9} is reminiscent of the wave equation, but with a discretized double derivative $\Delta_{n}$ and a $k-$dependent velocity of propagation:
\bea
\left[\frac{1}{4 \cos^2(k)}\frac{\partial^2 }{\partial t^2} - \Delta_{n}\right] \hat W(n,k;t) = 0.
\label{13}
\eea
The differential equation for the Wigner function at time $t$ is obtained by computing the expectation value of the expression above ($\psi$ here is static):
\bea
\left[\frac{1}{4 \cos^2(k)}\frac{\partial^2 }{\partial t^2} - \Delta_{n}\right] \< \hat W(n,k;t) \>_{\psi} = 0.
\label{13.1}
\eea
The solutions of this equation are superpositions of Bessel functions of any kind $\sum_m C_m J_{n+m}(4t\cos(k))$, as can be checked from the combination of known recurrences (10.6.1 and 10.6.2 in \cite{nist}). 
The coefficients $C_m$ are determined by the initial condition using $J_n(0) = \delta_{n,0}$. 
Now that we know that the evolution can be described with wave equations and no diffusion terms, we may infer the transformation of phase space variables by means of propagating pulses. In the continuous limit, $n \mapsto x$, the equation above becomes 
\bea
\left[\frac{1}{4 \cos^2(k)}\frac{\partial^2 }{\partial t^2} - \frac{\partial^2}{\partial x^2}\right] W(x,k;t) = 0,
\label{14}
\eea
 whose solutions correspond to the transformation
\bea
x(t) = x \pm 2\cos(k) t,\quad k(t) = k(0) = k
\label{15}
\eea
i.e. a shear transformation in $x,k$ with a modified velocity $2\cos(k)$. This satisfying result can be checked also by solving the Heisenberg equations of motion for $T, T^{\dagger}, N$:
\begin{equation}
N(t) = N(0) + t \left[T(0)+T^{\dagger}(0)\right],\quad T(t) = T(0).
\label{16}
\end{equation}
The behavior for an initial delta distribution in phase space is predicted by Eq. \eqref{15} and shown in Fig. \ref{wigfig1}. 
Further examples are given in the next subsection. 

\subsection{Examples in phase space}

\begin{figure}
\begin{picture}(100,100)(70,0)
\put(0,0){\includegraphics[width=.24\textwidth]{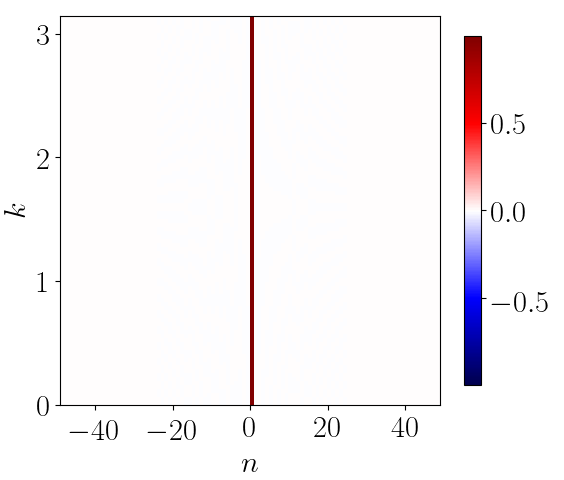}}
\put(125,0){\includegraphics[width=.24\textwidth]{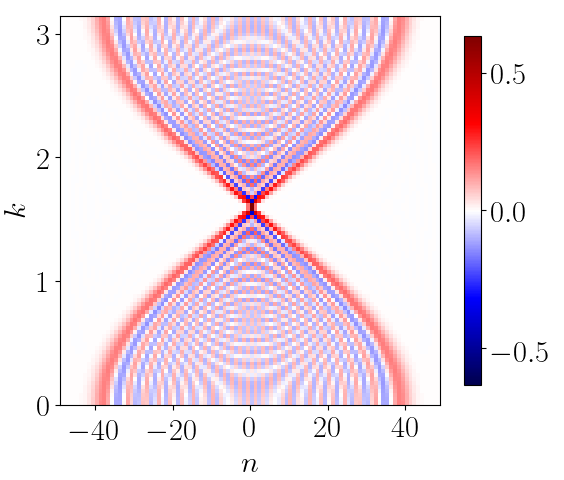}}
\put(15,87){(a)}
\put(140,87){(b)}
\end{picture}\\
\begin{picture}(100,100)(70,10)
\put(0,0){\includegraphics[width=.24\textwidth]{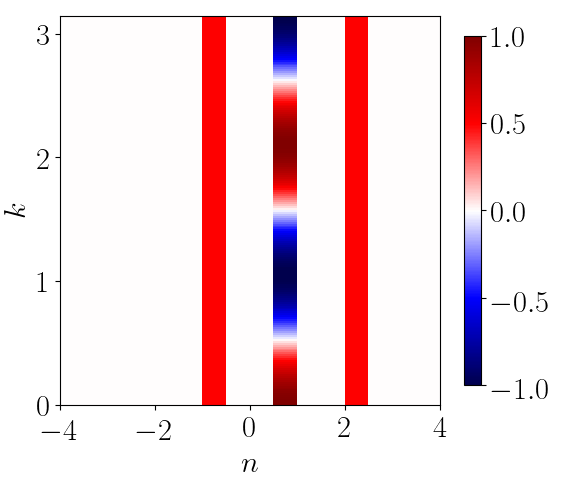}}
\put(125,0){\includegraphics[width=.24\textwidth]{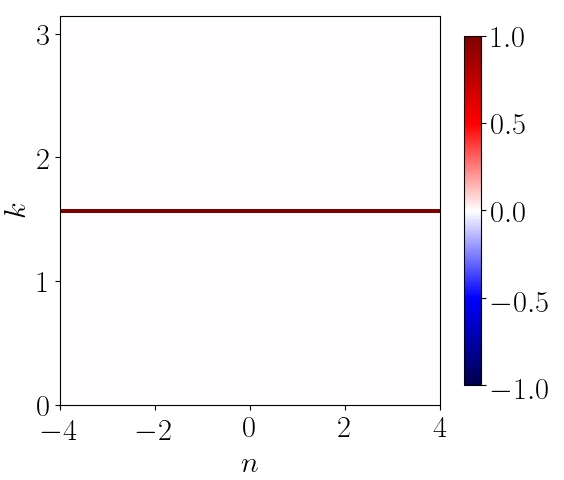}}
\put(15,87){(c)}
\put(140,87){(d)}
\end{picture}
  \caption{\label{wigfig1}
 Wigner function in the phase space of an infinite lattice with position $n$ and
 quasi-momentum $k$ for the following states: (a) A position eigenstate  $\ket n$ with $n=0$. (b) The same state after the evolution for a time $t=20$ under the influence of the Hamiltonian 
 in Eq. \eqref{8}. (c)
 A superpositon state $(\ket{n_1}+\ket{n_2})/\sqrt2$ with $n_1=-1$ and $n_2=2$. (d)
 A  quasi-momentum eigenstate $\ket k$ with $k=\pi/2$. 
  }
\end{figure}
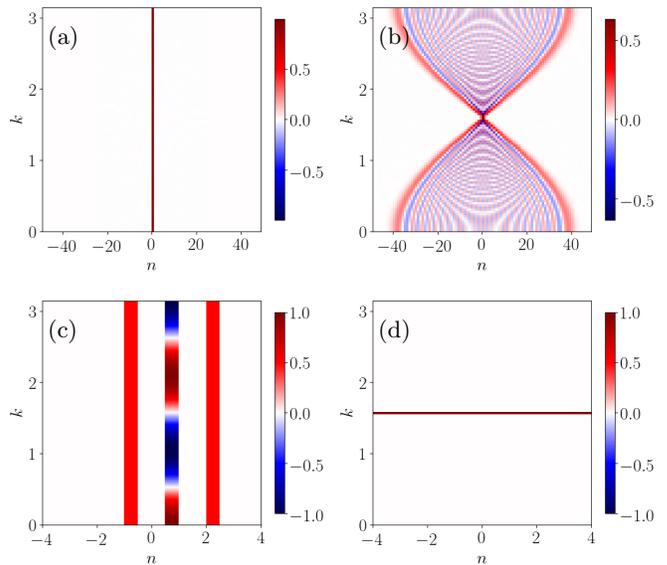

The Wigner function of position eigenstates can be easily computed from the definition
in Eq. \eqref{5}. One can check that for $\ket{n_1}$, one obtains
$W_{n_1}(n,k)=\delta_{n,n_1}/2\pi$, which simply represents a vertical stripe in phase space
as depicted in Fig. \ref{wigfig1} (a). 
%The time-evolution under the dynamics of the Hamiltonian
%in \eqref{8} is shown also in Fig. \ref{wigfig1}. 
This contrasts with the 
definition of \cite{hinarejos}, where the stripe is located at twice 
 the position value. The time evolution of such initial state under the
influence of the Hamiltonian in Eq. \eqref{8} is shown in Fig. \eqref{wigfig1} (b), which is plotted
for a time $t=20$. One can note that for the horizontal line $k=\pi/2$, the Wigner function
extends to values of $n=\pm 40$, well in accordance with Eq. \eqref{15}. The node at $k=0$ is also displayed. 

The case of a quasi-momentum eigenstate can also be 
evaluated, in this case for the state vector 
$\ket{k_1}$ the Wigner function is given by $W_{k_1}(n,k)=\delta(k-k_1)/2\pi$, i.e., 
a delta function in the quasi-momentum $k$. This
is due to the fact that one has a continuum of quasi-momentum states. As $\ket k$ is an eigenstate 
of the Hamiltonian  \eqref{8} it remains unchanged under its influence. The Wigner function for the
case of $k=\pi/2$ is shown in Fig \ref{wigfig1} (d).

Another interesting example is the superposition of two position eigenstates
$(\ket{n_1}+\ket{n_2})/\sqrt2$. For these type of states, the Wigner function
takes the form
\begin{equation}
  W_{n_1,n_2}(n,k)=\frac{\delta_{n,n_1}+\delta_{n,n_2}}{4\pi}+
 \delta_{n,\frac{n_1+n_2}{2}}
  \frac{
  \cos k(n_1-n_2)
  }{2\pi}.
 \label{catn}
\end{equation}
It is to be noted here the existence of interference fringes that arise exactly in
between the two vertical stripes representing the position states. The middle stripe
oscillates and, depending the values of $n_1$ and $n_2$, it can be located
at half integer values. In \ref{wigfig1} (c) we show an example with $n_1=-1$
and $n_2=2$.

The analogue in quasi-momentum space is given by the superposition
$(\ket{k_1}+\ket{k_2})/\sqrt2$. For this type of states, the Wigner function
is  given by
\begin{align}
  \label{catk}
  W_{k_1,k_2}(n,k)&=\frac{\delta(k-k_1)+\delta(k-k_2)}{4\pi}
  \\&+
  \delta\left(k-\frac{k_1+k_2}{2}\right)\frac{\cos [n(k_1-k_2)]}{2\pi}.
  \nonumber
\end{align}
In time, the position of the horizontal lines remains unchanged and only the central stripe will display an interference pattern 
shifted in time with velocity $v=2(\sin k_1-\sin k_2)/(k_1-k_2)$.

\section{Propagator of the discrete Wigner function}
\label{Propagator}
In this section we consider the dynamics of any system with discrete spectrum unbounded from above and
below, i.e., a spectrum that can be labeled by $n\in\mathbb Z$. This general construction will allow us
to treat different models later on. With the aforementioned assumptions, the 
Hamiltonian can be written in diagonal form in its eigenbasis as  
\bea
H =\sum_{n=-\infty}^{\infty}\varepsilon_{n} | n \>\<  n |. 
\label{19}
\eea
Here we have chosen to denote the eigenvectors of the Hamiltonian as $\ket n$ and
we will choose this basis to construct the corresponding Wigner function. The Hamiltonian  $H$ can also be 
expressed as a function of the position operator $N$. Therefore, 
a simple differential equation for $\hat{W}$ can be obtained by using the algebraic properties of commutators:
\begin{align}
\frac{d \hat{W}(n,k; t) }{d t} = i\left[H(N) -H(2n-N) \right] \hat{W}(n, k; t),
\label{25}
\end{align}
where it should be noted that neither $n$ nor $k$ depend on time so we use the total derivative with respect to $t$.
%where we have taken into account that $H(N) \ket n = \varepsilon_{n} \ket n$.
In order to deal with the number dependence of $H$, it is even more rewarding to compute derivatives with respect to $k$:
\begin{align}
\frac{i}{2}\frac{\partial \hat{W}}{\partial k} = (-N + n)\hat{W} \quad \mbox{or} \quad 
N \hat{W} =\left(n- \frac{i}{2} \partial_{k} \right)\hat{W}.
\label{26}
\end{align}
%
%Since $H$ is a function of $N$, it can be expanded in powers of $i \partial_{\kappa}$ around its analyticity domain.
%\footnote{The branch cut in (\ref{18}) is not harmful, as long as $\omega$ or $\delta \neq 0$.}. 
In this way, the evolution of $\hat{W}$ can be directly obtained from the definition of the Heisenberg picture, together with the algebraic properties of Eqs. \eqref{relT}and \eqref{3.1}:
\bea
\hat{W}(n,k; t) &=& e^{i t \left[ H(N)-H(2n-N) \right]} \hat{W}(n,k; 0).
\label{27}
\eea
Computing the expectation value of this relation with $\psi$, together with (\ref{26}), yields the desired result:
\begin{equation}
W(n,k;t) = e^{i t \left[ H \left( n-\frac{i}{2} \partial_{k} \right)
-H \left( n+\frac{i}{2} \partial_{k}  \right) \right]} W(n,k;0).
\label{28}
\end{equation}
It is rewarding to decompose the initial condition in Fourier modes using a $\delta$ function of period $\pi$ 
(note that Bloch waves have periodicity $2\pi$, 
but the Wigner function in \eqref{1} has half the period) 
\begin{equation}
W(n,k;0) =  \int_{0}^{2\pi} dk' \sum_{n'=-\infty}^{\infty}\frac{e^{2i n'(k'-k)}}{2\pi}  W(n,k';0)
\label{29}
\end{equation}
with the aim of solving any initial data problem {\it linearly.\ }This will lead to a propagator for the Wigner function. Inserting (\ref{29}) back into (\ref{28}) yields
\begin{equation}
W(n,k; t) = \int_{0}^{2\pi} dk' K(n, k-k'; t) W(n,k'; 0)
\label{29.1}
\end{equation}
with 
\begin{equation}
K(n, k; t) = \frac{1}{2\pi} 
\sum_{n'=-\infty}^{\infty} 
e^{it\left[\varepsilon_{n-n'}-\varepsilon_{n+n'} \right] - i2n'k}.
\label{30}
\end{equation}
The propagator $K$ can be shown to be real, as expected from the properties of the Wigner function:
\begin{equation}
K(n, k; t) = \frac{1}{2\pi} + \frac{1}{\pi} 
\sum_{n'=1}^{\infty} \cos \left\{ t\left[\varepsilon_{n+n'}-\varepsilon_{n-n'} \right] + 2n'k \right\}.
\label{31}
\end{equation}
This propagator is the main result of this section. One can also check its validity by 
noting that in  the Heisenberg picture, the operator $\hat W$ can be easily evaluated
from Eq. \eqref{7} as
\begin{equation}
\hat W(n,k;t)=\sum_{n'} e^{it[\varepsilon_{n-n'}-\varepsilon_{n+n'}]-i2n'k}\ketbra{n-n'}{n+n'}.
\label{Wtseries}
\end{equation}
After taking the expectation value of \eqref{Wtseries} and using the completeness relation in \eqref{complete} one obtains
the same result as in Eq. \eqref{29.1}, by realizing that
\begin{equation}
  \Tr{W(n',k')\hat W(n,k;t)}=K(n,k-k',t)\delta_{n,n'}.
  \label{}
\end{equation}

The form of the kernel is valid for any Hamiltonian with unbounded spectrum above and below. We shall see that for an atom-photon model, our kernel (\ref{31}) can be related to incoherent oscillations. 
%as well as some special limits where each $\varepsilon_n$ can be expanded to second order in  $n$. 
The Jacobi theta function and the possibility of Gaussian factorization arise if the energy is Taylor-expanded in second order in $n\pm n'$.
%Our computational procedure should be compared with the standard theory in continuous variables [{\bf Falta}] for generic Hamiltonians [{\bf Falta}], where diffusive (and incredibly contrived) terms arise in the corresponding differential equations.

\subsection{Example: superposition of point-like stimuli}
We devote this subsection to study the evolution of superposition of position and quasi-momentum states, whose Wigner function is
written in Eqs. \eqref{catn} and \eqref{catk} respectively. The Kernel derived in this section is particularly helpful to find
the time evolution of quasi-momentum eigenstates. For the Wigner function in \eqref{catk} and with the aid of \eqref{29.1} one obtains
\begin{align}
  \label{catkt}
  W_{k_1,k_2}&(n,k;t)=\frac{K(n,k-k_1,t)+K(n,k-k_2,t)}{4\pi}
  \nonumber
  \\&+
  K\left(n,k-\frac{k_1+k_2}{2},t\right)\frac{\cos [n(k_1-k_2)]}{2\pi}.
\end{align}
The case of a superposition of lattice points is simpler to solve in this case. Here the vertical stripes remain constant in position with a
dynamic interference pattern moving in the $k$ direction with velocity $v=(\varepsilon_{n_1}-\varepsilon_{n_2})/(n_1-n_2)$.

\section{Atom-field interaction models: The polariton phase space}
\label{Polariton}
%****************************************************************
In this section we identify the dressed states of the Jaynes-Cummings (JC) model  (and other atom-field systems) as the
position states of the lattice. The connection seems quite natural, as the spectrum is divided into a negative and
a positive part. This type of states describe hybrid excitations arising from the combination of a photonic
and an electronic excitation, which are also known as polaritons in the literature \cite{hartmann2006,angelakis,restrepo,koch}. Therefore,
we will refer to this space as  polariton or dressed state phase space.

\subsection{The Jaynes-Cummings model}
First we map the JC model to an infinite lattice made of excitation numbers. We work with the
JC Hamiltonian 
\cite{jaynes,shore} describing the interaction between a two-level atom with one mode of the radiation field 
\begin{equation}
H = 
\omega (a^\dagger a 
+\sigma_+ \sigma_-)
+\delta \sigma_+ \sigma_-
+ g (\sigma_+ a + \sigma_- a^{\dagger}). 
  \label{HamJC}
\end{equation}
The eigenvalues of $H$ can be labeled by an integer $n$, namely $\varepsilon_{n=0}=0$ and
\begin{equation}
\varepsilon_{n} = 
\omega |n| +\frac{  \delta}{2} + \mbox{sgn}(n) \sqrt{ \left( \frac{ \delta}{2} \right)^2 + g^2 |n| },\quad n \neq 0,
\label{jcEn}
\end{equation}
with the dressed states (or polariton states) given by $\ket{0}=\ket{0,-}$ and
\begin{equation}
\ket n  =
\cos \theta_{n}  \ket{|n| -1,+}  + \sin \theta_{n} \ket{|n|,-}, \quad n\neq 0
\label{dressed}
\end{equation}
%   $\ket0=\ket{0,-}$
where we have employed the shorthand
\begin{equation}
  \theta_n=\arctan\left(\frac{\varepsilon_n-|n|\omega}{g\sqrt{|n|}}\right).
  \label{theta}
\end{equation}
Furthermore,  we have introduced the states
\begin{equation}
  \ket{n_{\rm ph},\pm}=\ket{n_{\rm ph}}_{\rm ph}\otimes\ket\pm_{\rm at},\quad n_{\rm ph}\in \mathbb{N}_0
  \label{naked}
\end{equation}
which represent 
$n_{\rm ph}$ photons in the mode and an excited or de-excited atom for
$\ket +_{\rm at}$ and $\ket -_{\rm at}$ respectively.

In this case, the points in the lattice correspond to dressed state of the Jaynes-Cummings model. In
phase space, those states are represented by vertical stripes as depicted before in Fig. 
\ref{wigfig1} (a). A  product state of the form \eqref{naked} can always be represented as the
sum of two dressed states $\ket{n}$ and its negative counterpart. Therefore, 
the corresponding Wigner function is always formed by two vertical stripes positioned at the $n$
and a central stripe at $n=0$ with the interference pattern oscillating with frequency $n$,
similar to the case depicted in Fig. \ref{wigfig1} (c).
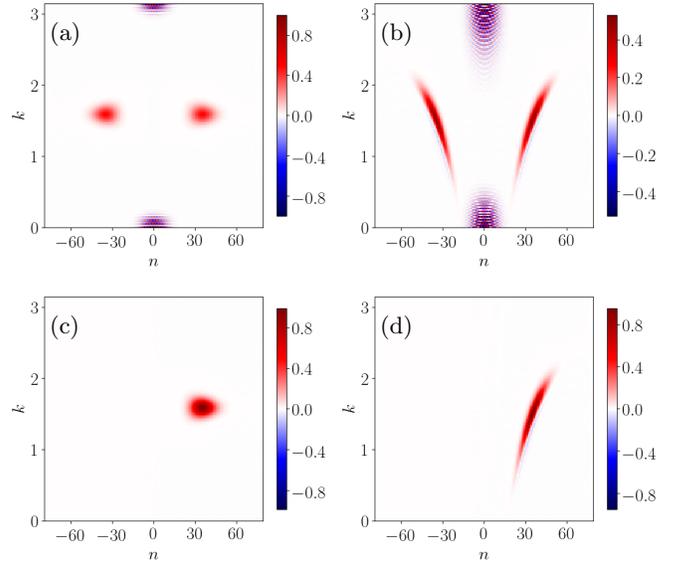
\begin{figure}
\begin{picture}(100,100)(70,0)
\put(0,0){\includegraphics[width=.24\textwidth]{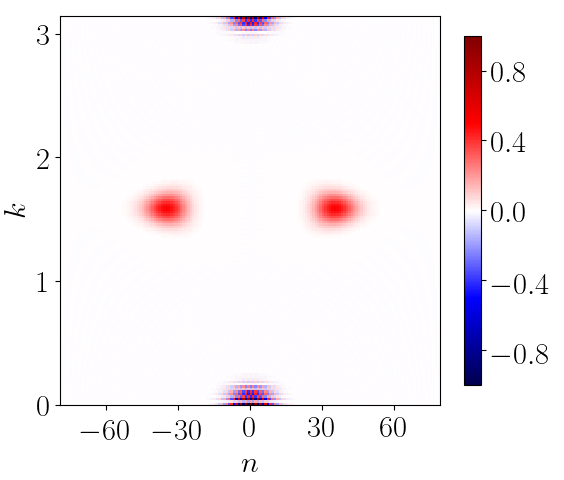}}
\put(125,0){\includegraphics[width=.24\textwidth]{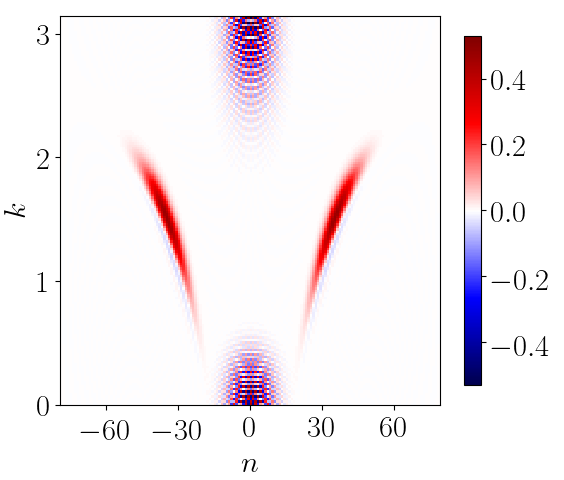}}
\put(15,87){(a)}
\put(140,87){(b)}
\end{picture}
\\
\begin{picture}(100,100)(70,10)
\put(0,0){\includegraphics[width=.24\textwidth]{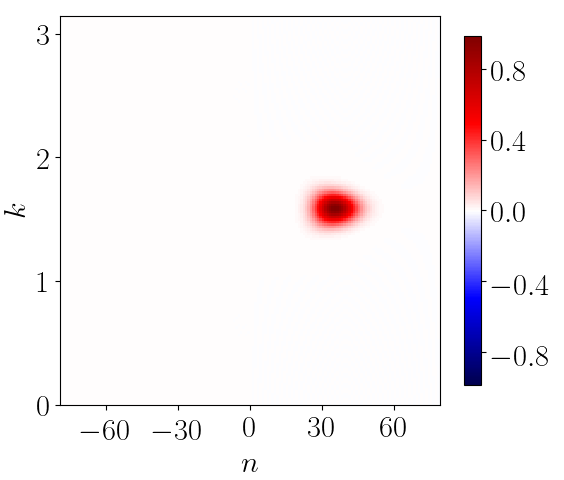}}
\put(125,0){\includegraphics[width=.24\textwidth]{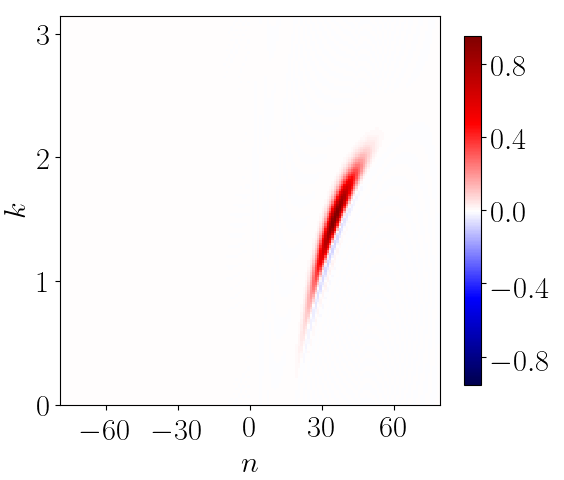}}
\put(15,87){(c)}
\put(140,87){(d)}
\end{picture}
  \caption{\label{wigfigcoh} 
  Wigner function in the polariton phase space of the Jaynes-Cummings model of: 
  (a) A coherent state of the light field and an atom in the ground state.
  (c) A dressed coherent state defined in \eqref{dressedcoh}. 
  (b) and (d) correspond to the time-evolved state, under the influence of the Hamiltonian in Eq. 
  \eqref{HamJC} for a revival time $t_r=2\pi|\alpha|/g$, of the initial state in (a) and (b) 
  respectively. In all cases $\alpha=6e^{i\pi/2}$.
  }
\end{figure}

A more familiar type of state, which is encountered in many quantum optical examples, is a product
state of an atomic state and a coherent state of the field. For the sake of simplicity, we will consider the dynamical behavior in the polariton phase space of a photonic coherent and a de-excited atom, namely the state
\begin{equation}
  \ket{\psi_\alpha}=\sum_{n_{\rm ph}=0}^\infty \frac{e^{-|\alpha|^2/2}\alpha^{n_{\rm ph}}}{\sqrt{n_{\rm ph}!}}\ket{n_{\rm ph},-}.
  \label{coherent}
\end{equation}
If the detuning is chosen to be zero, it is not hard to realize that this state takes the simple 
form
\begin{equation}
  \ket{\psi_\alpha}=\sum_{n=-\infty}^\infty 
  \frac{e^{-|\alpha|^2/2}\alpha^{|n|}}{\sqrt{2|n|!}}\ket{n},
  \label{coherent2}
\end{equation}
which leads to a Wigner function of the form
\begin{equation}
  W(n,k)=\sum_{n'}
  \frac{e^{-\frac{|\alpha|^2}{2}+i[2kn'+\phi(n_+'-n_-')]}|\alpha|^{n_+'+n_-'}}{\sqrt{4n_-'!n_+'!}}
  \label{wigcoh}
\end{equation}
with $n_\pm'=|n\pm n'|$. For $\alpha=0$ this expression reduces to $\delta_{n,0}$, as the state
with an empty cavity and the atom in the ground state is an eigenstate of the model. For
large values of $|\alpha|^2$, i.e., a large mean photon number, this Wigner function
has  three Gaussian peaks: two located at positions $n=\pm|\alpha|^2$ and a third one
at $n=0$ oscillating in $k$ with frequency $|\alpha|^2$. This can be realized by replacing
the Poissonian distributions in \eqref{wigcoh} by  Gaussian distributions and performing
the summation. A particular case for $|\alpha|^2=50$ is plotted in in Fig. \ref{wigfigcoh} with
its time-evolved plot at the revival time $t_r=2\pi|\alpha|/g$ \cite{schleich}.

One might argue that the product states we have considered so far, cannot be visualized in a simple
way in the polariton phase space. The reason is that we have not considered combined states 
of the atom-field system. This is the case where one can better exploit the phase space we have proposed. In this sense, dressed coherent states \cite{govia} are hybrid states of the atom-field system that
can be easily depicted in the polariton phase space.
In our notation they can be written as
\begin{equation}
  \ket{\alpha_-}=\sum_{n\in \mathbb{N}_0}
  \frac{e^{-|\alpha|^2/2}\alpha^{n}}{\sqrt{n!}}\ket{n}.
  \label{dressedcoh}
\end{equation}
Following the definition in \cite{govia}, a second coherent state $\ket{\alpha_+}$ can be
written as \eqref{dressedcoh} in our notation, but with the sum taken over $n<0$.
The state in Eq. \eqref{coherent} can be written as a sum of two of these
states. Its Wigner function appears as a Gaussian bump centered at $\pm|\alpha|^2$ in the polariton phase space as can be noted in Fig. \ref{wigfigcoh} (c) and its time evolution after a revival time in
Fig. \ref{wigfigcoh}~(d). It is worth commenting, that we have restricted our analysis of these
portraits to the definition in Eq. \eqref{1} using only integer values of $n$, i.e., we have 
disregarded the half-integer lattice and in consequence the interval $[\pi,2\pi)$ in the
quasi-momentum coordinate. The behaviour is quite similar and can be constructed from 
the data of our plotted functions.  In \cite{hinarejos} it was shown that these contributions 
display oscillations and have been named ghost images  \cite{arguelles}.

\subsubsection{Diffusion in phase eigenvalue}

The diffusion of quasi-momentum eigenstates $k=k_0$ can be obtained using (\ref{31}). The Wigner
function in this case is given by
\begin{align}
\label{39}
W(n,k;t) &= \frac{1}{2\pi}+  \\
&\frac{1}{\pi}\sum_{n'=1}^{\infty} \cos \left[ t \left( \varepsilon_{n+n'}-\varepsilon_{n-n'} \right)+2n'(k-k_0) \right].
\nonumber
\end{align}
When $t \rightarrow \infty$, very strong oscillations cancel out the terms in the sum and the limit of the diffusion is $W \rightarrow 1/2\pi$. The transient effects can be estimated by replacing the sum over $n'$ by an integral over $x$; the stationary phase approximation yields the following caustics  
\begin{align}
&\Phi(x,n, k;t) = t(\varepsilon_{n+x }-\varepsilon_{n-x}) + 2x(k-k_0), 
\nonumber \\ 
&k-k_0 = x^* t\frac{\sqrt{(\delta/2)^2 +g^2|x^*+n|}-\sqrt{(\delta/2)^2 +g^2|x^*-n|}}{\sqrt{\left[(\delta/2)^2 
+g^2|x^*+n|\right]\left[(\delta/2)^2 +g^2|x^*-n|\right]}}
\nonumber \\
&\delta_x\left[ \Phi \right]|_{x=x^*} = 0. 
\label{40}
\end{align}
The family of trajectories in phase space are obtained by replacing $x^* = x(n,k-k_0;t)$ back in $\Phi$. Maxima and minima of $W$ appear according to
\bea
\Phi(x^*, n, k; t)= 
\begin{cases}
  2q\pi & \text{for maxima} \\ 
  (2q+1)\pi & \text{for minima}. 
\end{cases}
\label{41}
\eea
A simplified form of the trajectories in phase space is obtained in the strong coupling limit $g \gg \delta$, where
\bea
x^* \approx \frac{(k-k_0)\delta}{2 t n g }.
\label{42}
\eea
The relation between $n, k, t$ is
\bea
\frac{(k-k_0)^2\delta}{ t n g } + \frac{2tn g}{\delta} = m \pi, \quad m\in \mathbb{Z}.
\label{43}
\eea
Although the Kernel can not be obtained in closed form for the Jaynes-Cummings system, due to the form of the eigenfrequencies, there are a few 
atom-interaction models that exhibit linear spectrum and therefore allow exact evaluation of $K$. We consider some of  them in
a subsequent subsection,  but before that, let us consider the diffusion process arising from the
inclusion of counter-rotating terms, i.e., the full atom-field interaction model. 

\subsection{The Rabi model}
In this  subsection, we consider the Rabi model, i.e., the Jaynes-Cummings model (JCM)
without the rotating-wave approximation. Although, the model has been shown to be integrable
\cite{braak} with a countable spectrum, here we do not define the polariton space in term
of these energies. In this case, we rather study the dynamics of the Rabi model
in the polariton phase space induced by the JCM. The Hamiltonian can be written as
\begin{equation}
  H=\omega a^\dagger a + \Omega \sigma_+\sigma_-+g(\sigma_++\sigma_-)(a^\dagger +a).
  \label{}
\end{equation}
The difference with the JCM is the term $a^\dagger \sigma_++a\sigma_-$, which describes
processes of simultaneous creation of a photon and excitation of the atom, and 
simultaneous annihilation and de-excitation. 

\begin{figure}
\begin{picture}(100,100)(70,0)
\put(0,0){\includegraphics[width=.24\textwidth]{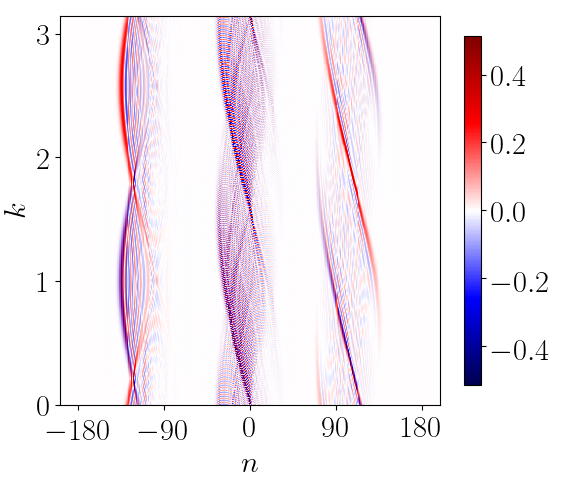}}
\put(125,0){\includegraphics[width=.24\textwidth]{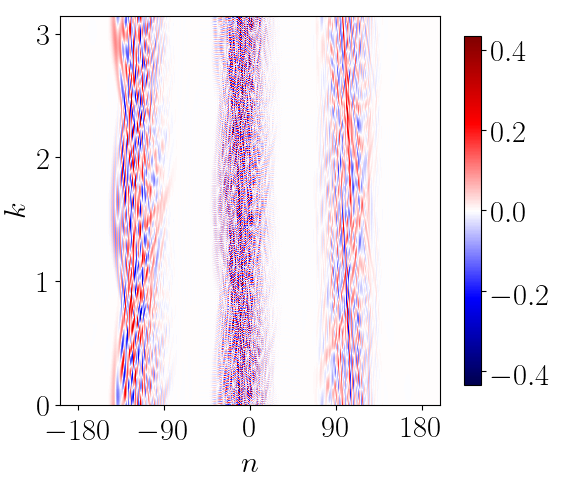}}
\put(15,87){(a)}
\put(140,87){(b)}
\end{picture}
\\
\begin{picture}(100,100)(70,10)
\put(0,0){\includegraphics[width=.24\textwidth]{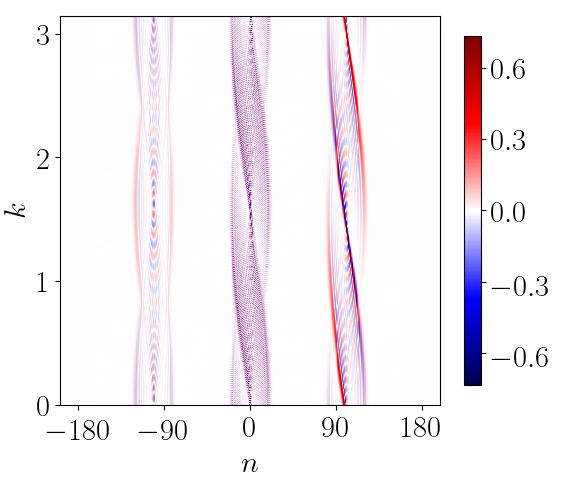}}
\put(125,0){\includegraphics[width=.24\textwidth]{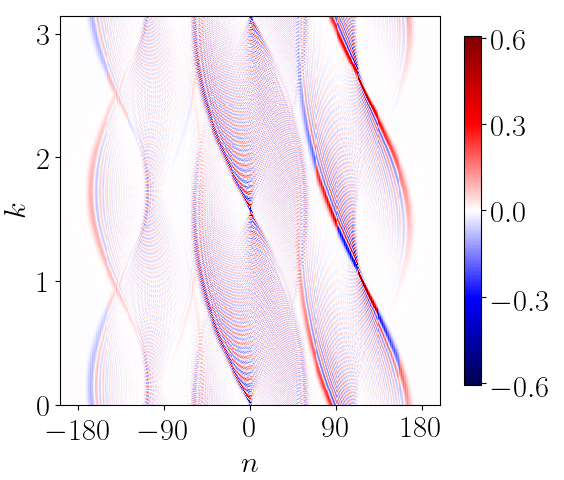}}
\put(15,87){(c)}
\put(140,87){(d)}
\end{picture}
  \caption{\label{wigfigrabi}
  Wigner function portraits  in the polariton phase space of the Jaynes-Cummings model 
  for an initial eigenstate $\ket {n=100}$ of the JCM and 
  after the evolution under the influence of the Rabi Hamiltonian.
  Moderate coupling $g/\omega=10$ and $\delta/\omega=1.5$ is considered with an interaction 
  times of (a) $t=2/g$ and (b) $t=2000/g$. In (c) $t=0.1/g$ and (b) $t=0.3/g$ we have
considered  strong coupling  $g/\omega=10$ and $\delta/\omega=1.5$.
  }
\end{figure}
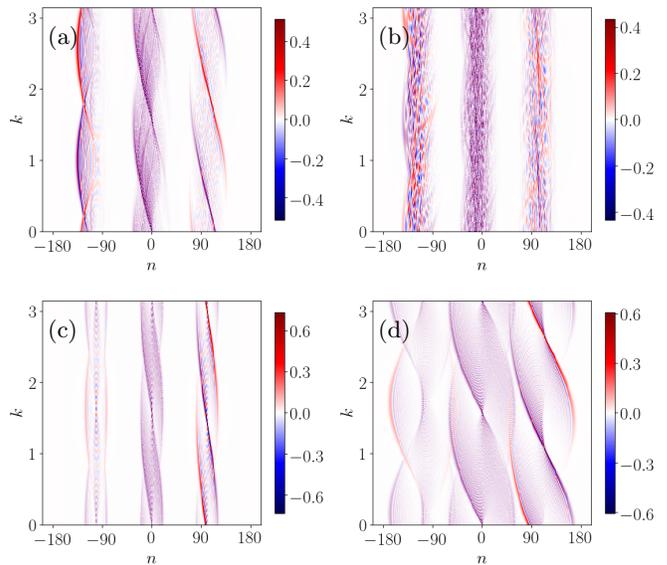

In fig. \ref{wigfigrabi} (a) and (b) we present the Wigner functions of an initial eigenstate of the
JC model $\ket{n=100}$ for an interaction time with the Rabi Hamiltonian of $t=2/g$ and
$t=2000/g$ respectively for a coupling strength of $g/\omega=1$. It can be noted
that even after a long interaction time, the excitations remain bounded and close to the initial
values, which is a consequence of a moderate choice of interaction strength, where
the rotating wave approximation is no longer valid, but the diffusion is still contained. 
Analogue portraits are shown  in Fig. \ref{wigfigrabi} (c) and (d) for a coupling strength ten times
larger and for interaction times of $t=0.1/g$ and
$t=0.3/g$ respectively. In this case,  the interaction is stronger and therefore one gets
faster diffusion in excitations. In both cases, it is interesting to realize that the behavior of the
diffusion resembles the braid-like patterns obtained in our tight binding model shown in Fig. 
\ref{wigfig1}~(b).

\subsection{The Buck-Sukumar model}
The Buck-Sukumar model \cite{buck} was motivated in the context of quantum optics as a simple atom-field interaction system with linear energies
and therefore perfect revivals of Rabi oscillations. The Hamiltonian 
at atomic resonance $\delta=0$ is given by
\begin{equation}
  H=
\omega (a^\dagger a 
+\sigma_+ \sigma_-)
%+\delta \sigma_+ \sigma_-
+  g\left(\sigma_- A^\dagger +\sigma_-A\right),
  \label{}
\end{equation}
with the ladder operator given by $A=a\sqrt{a^\dagger a}$.
The eigenvalues are once again linear in $n$ and can be written as $\varepsilon_{n=0}=0$ and
\begin{equation}
\varepsilon_{n} = 
\omega |n|  + g n,\quad n \neq 0,
\label{BsEn}
\end{equation}
with the eigenstates given by $\ket{0}=\ket{0,-}$ and
\begin{equation}
\ket n  =
\cos \varphi_{n}  \ket{|n| -1,+}  + \sin \varphi_{n} \ket{|n|,-}, \quad n\neq 0.
\label{dressedBs}
\end{equation}
Here one has to use the abbreviation 
\begin{equation}
  \varphi_n=\arctan\left(\frac{\varepsilon_n-|n|\omega}{g|n|}\right).
  \label{thetaBs}
\end{equation}
In terms of our propagator, this can be discussed since the sum can be performed in closed form.  
From Eq. \eqref{BsEn} one can infer that the sums will have two different fundamental frequencies $2\omega\pm2g$
which amounts to a superposition of delta functions and geometric sums. This implies quasiperiodicity in terms of the
two aforementioned fundamental frequencies. This will be valid for the evolution of any initial condition.  

\subsection{Far off resonant JC model}
A simplified version of the Jaynes-Cummings model can be obtained in the far off resonant limit, i.e.,  $\delta \gg g$. If this is
the case, one can derive an effective Hamiltonian  \cite{schleich,govia} of the form
\begin{equation}
H = 
\omega a^\dagger a 
+\omega' \sigma_+\sigma_-
- \frac{g^2}{\delta} \sigma_z a^{\dagger}a.
  \label{HJCBigD}
\end{equation}
with $\omega'=\omega+\delta- g^2/\delta$. 
A major advantage here is that the eigenvalues of $H$ are linear in $n$ and  take the form
$\varepsilon_{n=0}=0$ and 
\begin{equation}
\varepsilon_{n} =\omega|n|
+\frac{1-{\rm sgn}(n)}{2}\omega'
+\frac{g^2}{\delta} n
,\quad n \neq 0.
\end{equation}
The eigenvectors are simply given by
\begin{equation}
\ket n  = \ket{- n,+}, \quad n< 0; \quad
\ket n  = \ket{ n,-}, \quad n> 0,
\end{equation}
where we have expressed them in terms of the bare states defined in \eqref{naked}.

In terms of the Kernel for this example,  one can formulate a similar discussion as in the previous subsection. 
Here one also obtains a finite number of delta functions and geometric series. The fundamental frequencies  
are $g^2/\delta\pm\omega$ and $\omega'$.

\section{Conclusions}
 \label{Conclusion} 
 We have shown that discrete Wigner functions play an important role in visualizing the dynamics, not only in cavity-QED systems, but also
tight binding models such as crystals with evanescent transport and spin chains. For the latter case,
we have found a dynamical equation for the Wigner function in
the semi-discrete phase space, whose continuum limit corresponds to the wave equation. 

We have mapped the eigenstates of the Jaynes-Cummings model into this lattice and 
studied the Wigner function of relevant states of the field, including its dynamical evolution.
Furthermore, we have considered the diffusion process in this phase space under the influence
of counter-rotating terms of the atom-field interaction. 
 Although the Rabi model is integrable \cite{braak}, it displays rich enough dynamics so as to  violate 
 the conservation of excitation number  which has been essential in explaining, at the 
  textbook-level, the emission and absorption processes. 
  It should be noted that such a conservation law
  has been  utterly important in the derivation of Markovian master equations describing
  these phenomena in the context of open quantum systems \cite{carmichael}. However,
  under certain conditions, a Markovian description might be also too restrictive, therefore
  it is indispensable to provide appropriate tools in the study of more realistic situations. 
  Moreover we have  described the details of the dynamics by a careful computation of 
  quasi-distributions in the phase space of combined excitations. 
  The dynamical behavior of such excitations provides a better intuition on the diffusive 
  process of atom-field systems. As to Feynman's original question, we are sure that
  in this work we have provided a suitable framework to extend the discussion beyond the usual 
  conservation law.

\begin{acknowledgments}
  JMT acknowledges financial support by CONACyT through the program Repatriaciones 2016.
\end{acknowledgments}

\appendix

%\setcounter{section}{0}

%\section{Properties of discrete Wigner functions \label{sec:A1}}
%\section{\hspace{3cm} Derivation of discrete wave equations for Wigner functions}

\section{Computational details of marginals and Fourier transforms}
\label{ApComp}
The marginals of $W$ obtained by integrating over $k$ are trivial. On the other hand, the sums over $m$ require the following treatment
\begin{equation}
\sum_{m,n} \psi_{m+n} \psi^*_{m-n} e^{-2ikn} = \sum_{m,n}  \psi_{2m - n} \psi^*_{n} e^{2ik(n-m)}. 
\label{a1}
\end{equation}
Performing the sum over $m$ first with a change of index yields
\begin{align}
\sum_m \psi_{2m-n} e^{-2ikm}=
\begin{cases}
  \sum_{\mu \, \rm{odd}} \psi_{\mu} e^{-ik(\mu+n)},&  $n$ \text{ odd}\\
  \sum_{\mu \, \rm{even}} \psi_{\mu} e^{-ik(\mu+n)},& $n$ \text{ even}
\end{cases}
\label{a2}
\end{align}
and with this we obtain two contributions in (\ref{a1}) for $n$ odd and $n$ even:
\begin{align}
\sum_{\mu \, \rm{odd}} \psi_{\mu} e^{-ik\mu} \sum_{n \, \rm{odd}} \psi^*_{n} e^{ikn} +  \sum_{\mu \, \rm{even}} \psi_{\mu} e^{-ik\mu} 
\sum_{n \, \rm{even}} \psi^*_{n} e^{ikn}.
\label{a3}
\end{align}
These sums can be evaluated in terms of the discrete (full) Fourier transform
\bea
\tilde \psi_k = \sum_{\mu = -\infty}^{\infty} \psi_{\mu} e^{-ik \mu},
\label{a4}
\eea
by means of the summation formula $\sum_{n=-\infty}^{\infty} e^{-ink}=2\pi \sum_{r=-\infty}^{\infty} \delta(k + 2\pi r)$; we have three terms $r=-1,0,+1$ contributing in the second line of the following expression
\bea
\label{a5}
\sum_{\mu \, \rm{odd}} \psi_{\mu}e^{-ik \mu} &=& \frac{1}{2\pi} \int_{0}^{2\pi} dq \tilde \psi_q \sum_{\nu=-\infty}^{\infty} e^{-i(k-q)(2\nu+1)} \\ &=& \frac{1}{2} \left( -\tilde \psi_{k-\pi} + \tilde \psi_k -\tilde \psi_{k+\pi}\right) = \frac{1}{2} \tilde \psi_k - \tilde \psi_{k+\pi}\nonumber.
\eea
Similarly, the even sum gives
\bea
\sum_{\mu \, \rm{even}} \psi_{\mu} e^{-ik \mu} = \frac{1}{2} \tilde \psi_k + \tilde \psi_{k+\pi}.
\label{a6}
\eea
Finally, we get the marginal in the form
\begin{align}
\sum_{m =-\infty}^{\infty} &W^{sp}(m,k) = \frac{A^{sp}+C^{sp} + \left( B^{sp} e^{ik} + \rm{c.c.} \right)}{4\pi}
|\tilde \psi_k|^2 \nonumber \\ 
+& \left[ A^{sp}+C^{sp} - \left( B^{sp} e^{ik} + \rm{c.c.} \right) \right] |\tilde \psi_{k+\pi}|^2/\pi
\label{a7}
\end{align}
With this expression we are ready to solve for the coefficients $A,B,C$ as discussed in the text.

\end{document}